\documentclass[12pt, onecolumn]{IEEEtran}
\usepackage{epsf}
\usepackage{graphicx}% Extended graphics
\usepackage{amsmath} \usepackage{amssymb}

\newtheorem{theorem}{Theorem}[section]

\long\def\symbolfootnote[#1]#2{\begingroup%
\def\thefootnote{\fnsymbol{footnote}}\footnote[#1]{#2}\endgroup}

 \def\DD{{\cal D}}

\def\dref#1{(\ref{#1})}

\def\be{\begin{equation}} \def\ee{\end{equation}}
\def\ba{\begin{array}} \def\ea{\end{array}} \def\bna{\begin{eqnarray}}
\def\ena{\end{eqnarray}}

 \def\NN{{\cal N}}

\def\DD{{\cal D}}
\def\SS{{\cal S}}
 \def\XX{{\cal X}}   
\def\DD{{\cal D}}

\def\YY{{\cal Y}}

 \def\bna{\begin{eqnarray}}
\def\ena{\end{eqnarray}} \def\dref#1{(\ref{#1})}
\begin{document}
\title{An Improvement of Cover/El Gamal's Compress-and-Forward Relay Scheme}

\author{\authorblockN{Liang-Liang Xie} \\
\authorblockA{\small Department of Electrical and Computer Engineering\\
University of Waterloo, Waterloo, ON, Canada N2L 3G1 \\
Email: llxie@ece.uwaterloo.ca} }

\maketitle

\begin{abstract}

The compress-and-forward relay scheme developed by (Cover and El Gamal, 1979) is improved with a modification on the decoding process. The improvement follows as a result of realizing that it is not necessary for the destination to decode the compressed observation of the relay; and even if the compressed observation is to be decoded, it can be more easily done by joint decoding with the original message, rather than in a successive way. An extension to multiple relays is also discussed.

\end{abstract}

\section{Introduction}

\begin{figure}[hbt]
\centering
\includegraphics[width=2.6in]{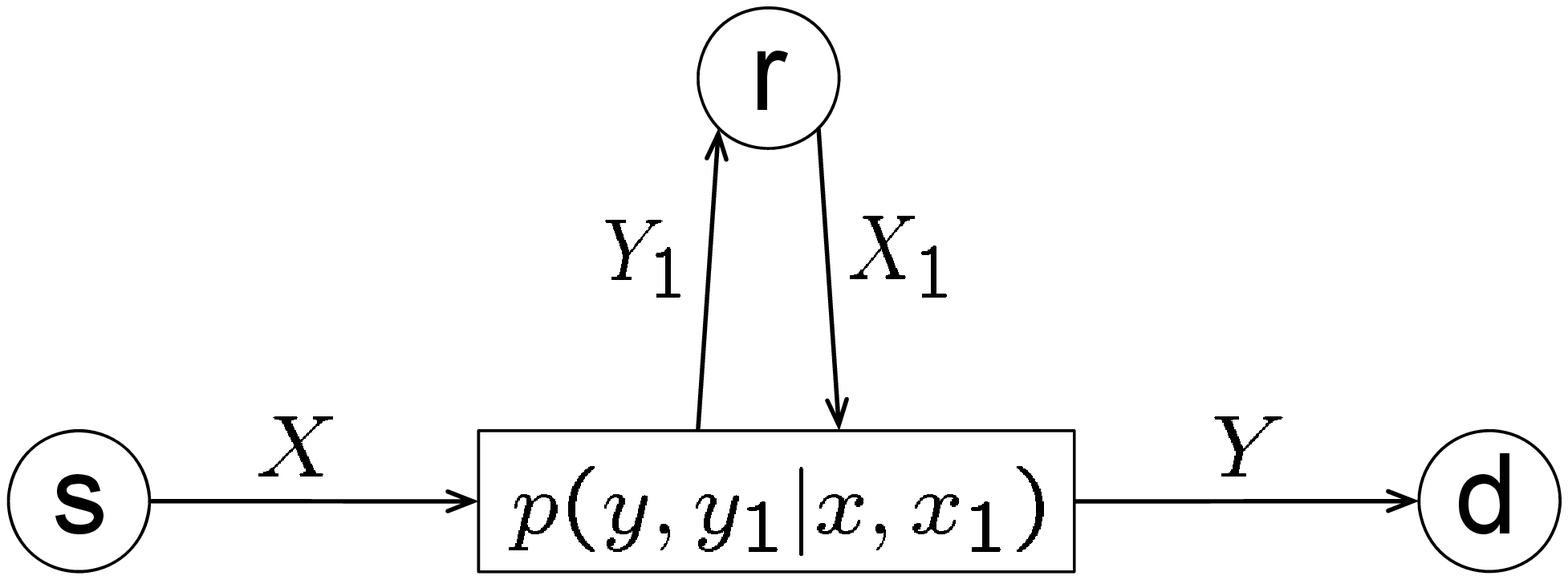}
\caption{The relay channel.}
 \label{fig1}
\end{figure}

The relay channel, originally proposed in \cite{van71}, models a communication scenario where there is a relay node that can help the information transmission between the source and the destination, as shown in Fig. \ref{fig1}. Two fundamentally different relay strategies were developed in \cite{covelg79}, which, depending on whether the relay decodes the information or not, are generally known as {\it decode-and-forward} and {\it compress-and-forward} respectively. The compress-and-forward relay strategy is used when the relay cannot decode the message sent by the source, but still can help by compressing and forwarding its observation to the destination.

In the compress-and-forward coding scheme developed in \cite{covelg79}, the relay first compresses its observation $Y_1$ into $\hat Y_1$, and then forwards this compressed version to the destination via $X_1$. This compression is generally necessary since the destination may not be able to completely recover $Y_1$. Instead, the compressed version $\hat Y_1$ can be recovered, as long as the following constraint is satisfied:
\begin{equation}
\label{constraint}
I(X_1;Y)>I(Y_1;\hat Y_1|X_1,Y).
\end{equation}
Then, based on $\hat Y_1$ and $Y$, the destination can decode the original message $X$ if the rate
\begin{equation}
\label{cov}
R<I(X;\hat Y_1,Y|X_1).
\end{equation}

In this paper, we propose a modification of this compress-and-forward coding scheme by realizing that it is not necessary to recover $\hat Y_1$ since the original problem is to decode $X$ only; and even if $\hat Y_1$ is to be decoded, it can be done by jointly decoding $\hat Y_1$ and $X$, instead of successively decoding $\hat Y_1$ and then $X$.

We will show that without decoding $\hat Y_1$, the constraint \dref{constraint} is not needed, and the achievable rate is more generally given by
\begin{equation}
\label{rate1}
R<I(X;\hat Y_1,Y|X_1)-\max\{0,I(Y_1;\hat Y_1|X_1,Y)-I(X_1;Y)\}.
\end{equation}
Obviously, any rate satisfying \dref{constraint}-\dref{cov} also satisfies \dref{rate1}. However, it remains a question whether there are interesting channel models where \dref{rate1} is strictly larger than \dref{constraint}-\dref{cov}. This problem will not be addressed here. Instead, we point out an immediate advantage of \dref{rate1} over \dref{constraint}-\dref{cov}. For \dref{constraint}-\dref{cov}, the relay needs to know the value of $I(Y_1;\hat Y_1|X_1,Y)$ in order to decide on the appropriate compressed version $\hat Y_1$ to choose. This requires the knowledge of the channel dynamics from $X$ to $Y$, which may be difficult to obtain for the relay, e.g., in wireless communications. However, this is not necessary for \dref{rate1}, where the relay can choose any version $\hat Y_1$ that is sufficiently close to $Y_1$, since $\hat Y_1$ is not to be decoded.

What if we also want to decode $\hat Y_1$? It turns out that by jointly decoding $\hat Y_1$ and $X$, the constraint \dref{constraint} is not necessary; instead, we need a less strict inequality as the following:
\begin{equation}
\label{constraint2}
I(X_1;Y)>I(Y_1;\hat Y_1|X_1,Y,X)
\end{equation}
where, obviously, the difference from \dref{constraint} is the additional information provided by $X$.

\section{The Single Relay Case}

Formally, the single-relay channel depicted in Fig. \ref{fig1} can be
denoted by
$$
(\XX\times \XX_1,\,
p(y,y_1|x,x_1),\,\YY \times \YY_1)
$$
where, $\XX$ and $\XX_1$
are the transmitter alphabets of the source and the relay respectively,
$\YY$ and $\YY_1$ are the receiver alphabets of the destination and
the relay respectively, and a collection of probability distributions
$p(\cdot,\cdot|x,x_1)$ on $\YY\times \YY_1$, one for each
$(x,x_1)\in \XX\times \XX_1$. The interpretation is that $x$
is the input to the channel from the source, $y$ is
the output of the channel to the destination, and $y_1$ is
the output received by the relay. The relay sends an input $x_1$ based on what it has received:
\begin{equation}
\label{processing}
x_1(t)=f_t(y_1(t-1),y_1(t-2),\ldots), \quad \mbox{ for every time } t,
\end{equation}
where $f_t(\cdot)$ can be any causal function. Note that a one-step
time delay is assumed in \dref{processing} to account for the
signal processing time at the relay.

\begin{theorem}
\label{th1}
For the single-relay channel depicted in Fig. \ref{fig1}, by the modified compress-and-forward coding scheme, a rate $R$ is achievable if it satisfies
\begin{equation}
\label{eq1}
R<I(X;\hat Y_1,Y|X_1)-\max\{0,I(Y_1;\hat Y_1|X_1,Y)-I(X_1;Y)\}
\end{equation}
for some $p(x)p(x_1)p(\hat y_1|y_1,x_1)$. In addition, the compressed version $\hat Y_1$ can be decoded if
\begin{equation}
\label{eq2}
I(X_1;Y)>I(Y_1;\hat Y_1|X_1,Y,X).
\end{equation}
\end{theorem}

In the modified scheme, the codebook generation and encoding process is exactly the same as that in the proof of Theorem 6 of \cite{covelg79}. The modification is only on the decoding process at the destination: i) The destination finds the unique $X$ sequence that is jointly typical with the $Y$ sequence received, and also with a $\hat Y_1$ sequence from the specific bin sent by the relay via $X_1$; ii) If the $\hat Y_1$ sequence is to be decoded, the destination finds the unique pair of $X$ sequence and $\hat Y_1$ sequence from the specific bin that are jointly typical with the $Y$ sequence received.

\section{Extension to Multiple Relays}

An extension of Cover/El Gamal's compress-and-forward coding scheme to multiple relays was presented in \cite{kragasgup05}. We can also extend the modified scheme to multiple relays.

\begin{figure}[hbt]
\centering
\includegraphics[width=3.3in]{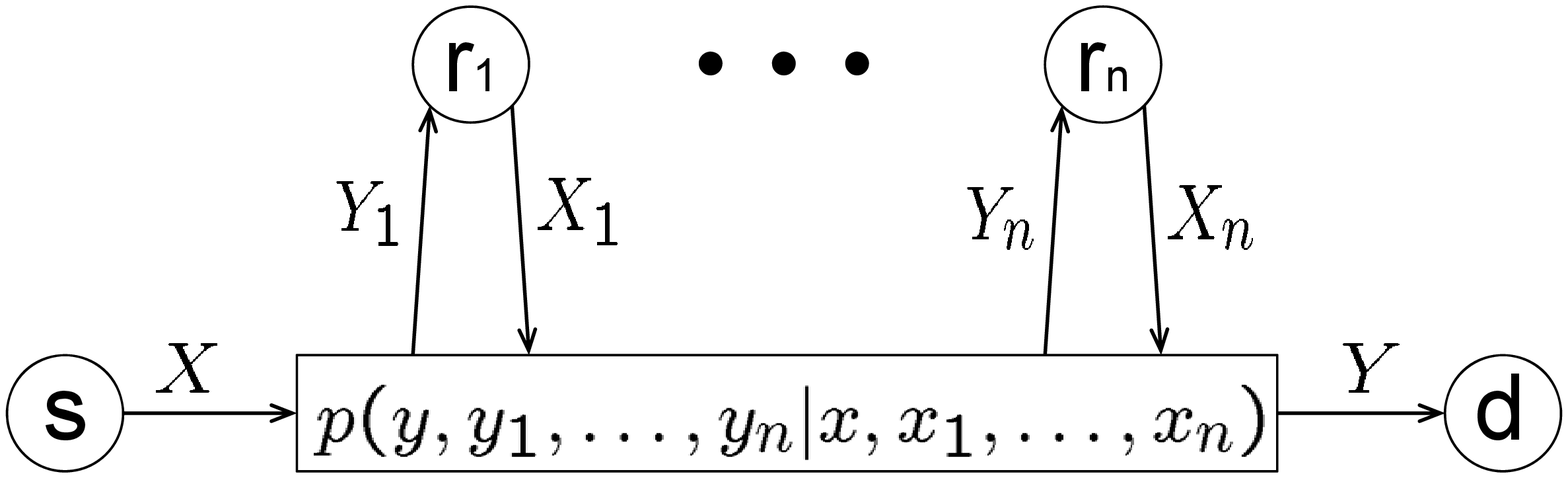}
\caption{A multiple-relay channel.}
 \label{fig2}
\end{figure}

A multiple-relay channel is depicted in Fig. \ref{fig2}, which can be
denoted by
$$
(\XX\times \XX_1\times\cdots\times \XX_n,\,
p(y,y_1,\ldots,y_n|x,x_1,\ldots,x_n),\,\YY \times \YY_1\times \cdots \times \YY_n)
$$
where, $\XX,\XX_1,\ldots,\XX_n$
are the transmitter alphabets of the source and the relays respectively,
$\YY,\YY_1,\ldots,\YY_n$ are the receiver alphabets of the destination and
the relays respectively, and a collection of probability distributions
$p(\cdot,\cdot,\ldots,\cdot|x,x_1,\ldots,x_n)$ on $\YY\times \YY_1\times \cdots \times \YY_n$, one for each
$(x,x_1,\ldots,x_n)\in \XX\times \XX_1\times\cdots\times \XX_n$. The interpretation is that $x$
is the input to the channel from the source, $y$ is
the output of the channel to the destination, and $y_i$ is
the output received by the $i$-th relay. The $i$-th relay sends an input $x_i$ based on what it has received:
\begin{equation}
\label{processing2}
x_i(t)=f_{i,t}(y_i(t-1),y_i(t-2),\ldots), \quad \mbox{ for every time } t,
\end{equation}
where $f_{i,t}(\cdot)$ can be any causal function.

Before presenting the achievability result, we introduce some simplified notations. Denote the set $\NN=\{1,2,\ldots,n\}$, and for any subset $\SS\subseteq \NN$, let $X_{\SS}=\{X_i,i\in \SS\}$, and use similar notations for other variables. We have the following achievability result.

\begin{theorem}
\label{th2}
For the multiple-relay channel depicted in Fig. \ref{fig2}, by the modified compress-and-forward coding scheme, a rate $R$ is achievable if for some
$$
p(x)p(x_1)\cdots p(x_n)p(\hat y_1|y_1,x_1)\cdots p(\hat y_n|y_n,x_n),
$$
there exists a rate vector $\{R_i,i=1,\ldots,n\}$ satisfying
\begin{equation}
\label{Rconstraint}
\sum_{i\in \SS_1}R_i<I(X_{\SS_1};Y|X_{\SS_1^c})
\end{equation}
for any subset ${{\cal S}_1}\subseteq {\cal N}$, such that for
any subset ${\cal S}\subseteq {\cal N}$,
\begin{equation}
\label{eq21}
R<I(X;\hat Y_\NN, Y|X_\NN)-H(\hat Y_\SS|\hat Y_{\SS^c},Y,X_\NN)+\sum_{i\in \SS}H(\hat Y_i|Y_i,X_i)+\sum_{i\in \SS}R_i.
\end{equation}
In addition, a subset of the compressed version $\hat Y_\DD$ for some $\DD\subseteq \NN$ can be decoded, if for any $\SS\subseteq \NN$ with $\SS\cap \DD\neq \emptyset$,
\begin{equation}
\label{eq22}
H(\hat Y_\SS|\hat Y_{\SS^c},Y,X,X_\NN)-\sum_{i\in \SS}H(\hat Y_i|Y_i,X_i)<\sum_{i\in \SS} R_i.
\end{equation}
\end{theorem}

It is easy to check that Theorem \ref{th2} implies Theorem \ref{th1}, by noting the Markov Chain $(X, Y)\rightarrow (X_1,Y_1) \rightarrow \hat Y_1$.

\section{Further Improvement}

Furthermore, we can even consider joint decoding with $X_{\NN}$. Then the constraint \dref{Rconstraint} is not necessary for the decoding of $X_\NN$, with the help of $X$ and $\hat Y_\NN$ from the previous block. For this, we have the following achievability result.

\begin{theorem}
\label{th3}
For the multiple-relay channel depicted in Fig. \ref{fig2}, a rate $R$ is achievable if for some
$$
p(x)p(x_1)\cdots p(x_n)p(\hat y_1|y_1,x_1)\cdots p(\hat y_n|y_n,x_n),
$$
there exists a rate vector $\{R_i,i=1,\ldots,n\}$ such that for
any ${\cal S}_1\subseteq {\cal S}\subseteq {\cal N}$,
\begin{equation}
\label{eq31}
R<I(X;\hat Y_\NN, Y|X_\NN)-H(\hat Y_\SS|\hat Y_{\SS^c},Y,X_\NN)+\sum_{i\in \SS}H(\hat Y_i|Y_i,X_i)+\sum_{i\in \SS\backslash \SS_1}R_i +I(X_{\SS_1};Y|X_{\SS_1^c})
\end{equation}
and
\begin{equation}
\label{eq31'}
H(\hat Y_\SS|\hat Y_{\SS^c},Y,X,X_\NN)-\sum_{i\in \SS}H(\hat Y_i|Y_i,X_i)-\sum_{i\in \SS\backslash \SS_1} R_i-I(X_{\SS_1};Y|X_{\SS_1^c})<0.
\end{equation}
In addition, a subset of the compressed version $\hat Y_\DD$ for some $\DD\subseteq \NN$ can be decoded, if for any $\SS\subseteq \NN$ with $\SS\cap \DD\neq \emptyset$,
\begin{equation}
\label{eq32}
H(\hat Y_\SS|\hat Y_{\SS^c},Y,X,X_\NN)-\sum_{i\in \SS}H(\hat Y_i|Y_i,X_i)<\sum_{i\in \SS} R_i.
\end{equation}
\end{theorem}

\end{document}